\documentclass[aps,prb,two column,superscriptaddress]{revtex4-1}

\usepackage{gensymb}
\usepackage{hyperref}
\usepackage{graphicx}
\usepackage{amsmath}
\usepackage{amssymb}
\usepackage{bm}
\usepackage{color}
\usepackage{bookmark}
\usepackage{tabularx}
\usepackage{mathtools}
\usepackage{microtype}
\usepackage{cancel}
\usepackage{relsize}
\usepackage{textcomp}
\usepackage{ulem}

\begin{document}
\newcommand{\Ss}{\textsubscript}
\newcommand{\Us}{\textsuperscript}
\newcommand{\STO}{SrTiO\Ss{3}}
\newcommand{\GTO}{GdTiO\Ss{3}}
\newcommand{\GTOx}{GdTiO\Ss{3+x}}
\newcommand{\LAO}{LaAlO\Ss{3}}
\newcommand{\LGAO}{La\Ss{1-x}Gd\Ss{x}AlO\Ss{3}}
\newcommand{\GLTO}{Gd\Ss{1-y}La\Ss{y}Ti\Ss{z}Al\Ss{1-z}O\Ss{3}}
\newcommand{\LTO}{LaTiO\Ss{3}}
\newcommand{\RTO}{RETiO\Ss{3}}
\newcommand{\NGO}{NdGaO\Ss{3}}
\newcommand{\LCO}{LaCrO\Ss{3}}
\newcommand{\PyroCl}{Re\Ss{2}Ti\Ss{2}O\Ss{7}}
\newcommand{\ETO}{EuTiO\Ss{3}}
\newcommand{\ox}{O\Ss{2}}
\newcommand{\tit}{Ti\Us{3+}}
\newcommand{\tif}{Ti\Us{4+}}
\newcommand{\DC}{\degree C}
\newcommand{\mx}{\times}
\newcommand{\rev}{\cite}
\newcommand{\RAN}{$R_{xy}^{AHE}$}
\newcommand{\dxy}{3d\Ss{xy}}
\newcommand{\dxyz}{3d\Ss{xz/yz}}
\newcommand{\dxyzo}{d\Ss{xz/yz}}
\newcommand{\dxyo}{d\Ss{xy}}

\title{Gate-tuned Anomalous Hall Effect Driven by Rashba Splitting in Intermixed \LAO/\GTO/\STO}


\author{N. Lebedev}
 \email[Corresponding author:]{lebedev@physics.leidenuniv.nl}
\affiliation{Kamerlingh Onnes Laboratory, Leiden University, P.O. Box 9504, 2300 RA Leiden, The Netherlands}
\author{M. Stehno}
\affiliation{Physikalisches Institut (EP 3), Universität Würzburg, Am Hubland 97074 Würzburg, Germany}
\author{A. Rana}
\affiliation{Center for Advanced Materials and Devices, BML Munjal University (Hero Group), Gurgaon, India - 122413}
\affiliation{MESA+ Institute for Nanotechnology, University of Twente, P.O. Box 217, 7500 AE Enschede, The Netherlands}
\author{P. Reith}
\affiliation{MESA+ Institute for Nanotechnology, University of Twente, P.O. Box 217, 7500 AE Enschede, The Netherlands}
\author{N. Gauquelin}
\author{J. Verbeeck}
\affiliation{Electron Microscopy for Materials Science, University of Antwerp, Campus Groenenborger Groenenborgerlaan 171, 2020 Antwerpen, Belgium}
\author{H. Hilgenkamp}
\author{A. Brinkman}
\affiliation{MESA+ Institute for Nanotechnology, University of Twente, P.O. Box 217, 7500 AE Enschede, The Netherlands}
\author{J. Aarts}
\affiliation{Kamerlingh Onnes Laboratory, Leiden University, P.O. Box 9504, 2300 RA Leiden, The Netherlands}

\date{\today}

\begin{abstract}
The Anomalous Hall Effect (AHE) is an important quantity in determining the properties and understanding the behavior of the two-dimensional electron system forming at the interface of \STO-based oxide heterostructures. The occurrence of AHE is often interpreted as a signature of ferromagnetism, but it is becoming more and more clear that also paramagnets may contribute to AHE. We studied the influence of magnetic ions by measuring intermixed \LAO/\GTO/\STO{} at temperatures below 10 K. We find that, as function of gate voltage, the system undergoes a Lifshitz transition, while at the same time an onset of AHE is observed. However, we do not observe clear signs of ferromagnetism. We argue the AHE to be due to the change in Rashba spin-orbit coupling at the Lifshitz transition and conclude that also paramagnetic moments which are easily polarizable at low temperatures and high magnetic fields lead to the presence of AHE, which needs to be taken into account when extracting carrier densities and mobilities.
\end{abstract}


\maketitle

\section{Introduction\label{intro}}
The two-dimensional electron system (2DES) which is present at \STO -based oxide interfaces is of interest as a model system for the physics of band formation and electrical transport in a quantum well where $3d$ electrons are the carriers. Moreover, the system has built-in electrical tunability, since the high permittivity of the \STO{} substrate allows it to be used as a back gate, thereby varying the shape of the well, the number of carriers, and the population of the various $3d$ subbands. One of the still outstanding questions is whether and how the 2DES can be used as a platform for spintronics, meaning that the electron system can be (tunably) magnetically polarized and furnish not only charge current but also spin currents. The occurrence of Rashba-type spin-orbit coupling (SOC) at the interface is of obvious importance here, and it should be noted that the effect of back-gating does not only change the carrier concentration at the interface, but also changes the strength of the SOC~\cite{BenShalomPRL,CavigliaPRL}. This is a somewhat subtle band structure effect in which the system switches from one- to two-carrier transport at the so-called Lifshitz point \cite{JoshuaNCOM}. That leads to a strong increase in SOC, to a change in coupling between itinerant electrons and localised moments and to occurrence of the Anomalous Hall Effect(AHE), even without the presence of magnetic ions in the system~\cite{JoshuaPNAS}. More recently the effect was utilized for spin-to-charge conversion~\cite{LesneNMAT,WangNL}. The AHE is often used to detect magnetism in a system, for instance as it arises  in the walls between the tetragonal domains in \STO~which form below 105~K \cite{ChrisNPhys}. Defect control via Sr-vacancies was also proposed as way to engineer magnetic polarization and magnetic ordering, and AHE was used as a tool to detect such polarization~\cite{GunkelPRX}. Still, for engineering magnetism, magnetic ions should be advantageous. Magnetic $4f$ ions can easily be substituted for La, while using titanates rather than \LAO{} yields (magnetic) Ti$^{3+}$ ions. In this spirit, delta doping was used as a tool to enhance the amount of magnetic ions at the interface by using an ultrathin layer of EuTiO$_3$~\cite{NMATLAOETO}. Still, the mechanisms which are responsible for the occurrence of (tunable) AHE in these structures, and the role of SOC as an ingredient, have not yet been completely understood. \\
\begin{figure*}
\includegraphics[scale=0.30]{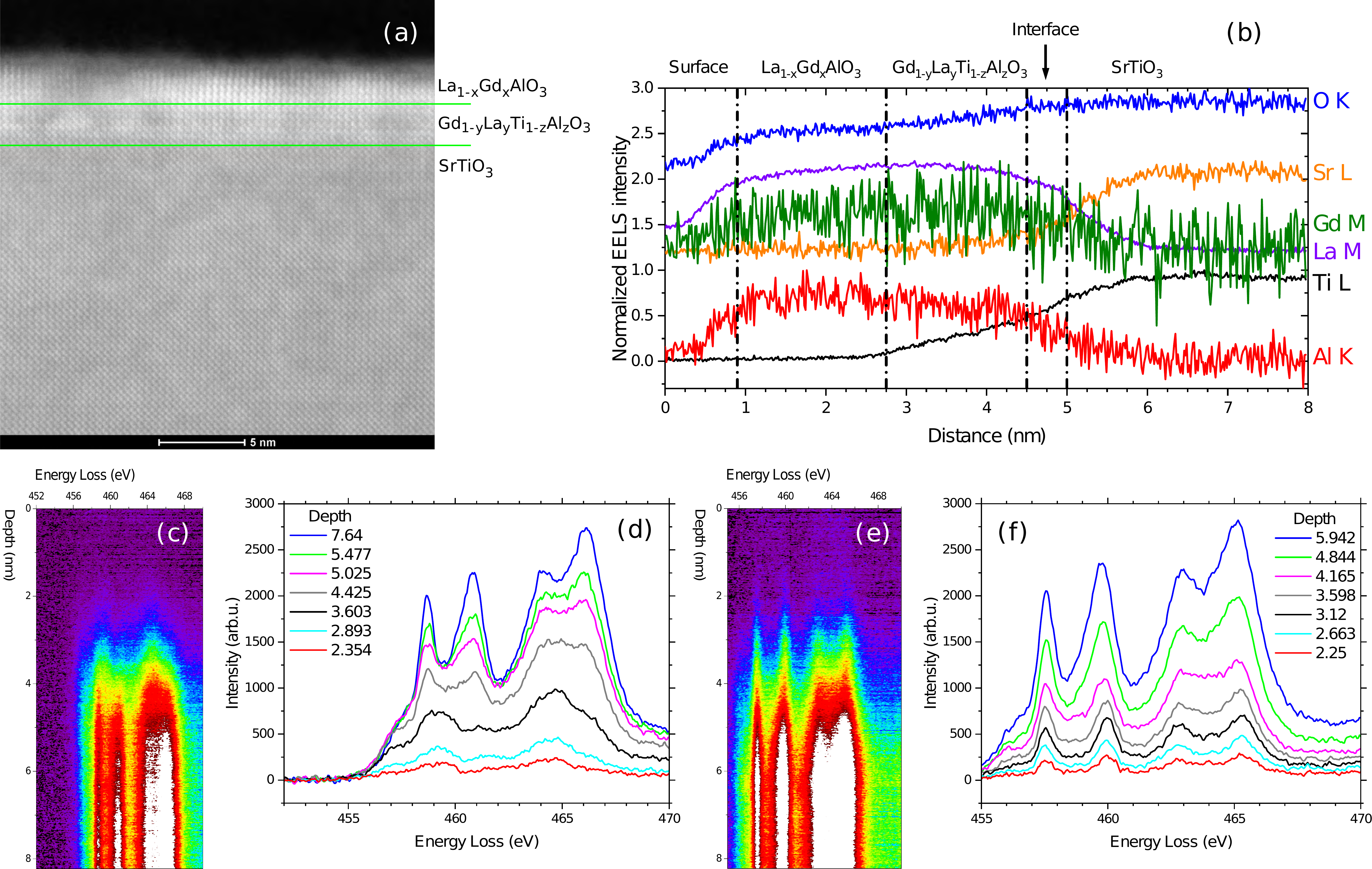}
\caption{(a) TEM image of crystal structure at and close to the interface. The green lines show the position of the interface with STO, mainly from the Sr EELS data; and the presumed interface between GTO and LAO, taken from the Ti EELS data. (b) EELS analysis of spatial distribution of the various elements. (c) Ti L edge and (d) corresponding spectra in region with presence of \tit{} in GLTAO layer.(e) Ti L edge and (f) corresponding spectra in region without of \tit{} in GLTAO layer.\label{Fig1}}
\end{figure*}
In our work we also implemented the delta doping method to study gate-tunable AHE. Previously Stornaiuolo {\it et al.} showed that by adding 2 unit cells of ferromagnetic \ETO{} (ETO) at the interface between the non-magnetic band insulators \LAO{} (LAO) and \STO{} (STO) magnetism and a tunable AHE can be realized\rev{NMATLAOETO}. ETO is a band insulator with Ti in the same \tif{} oxidation state as STO, meaning that ETO is non-polar along the (001) crystal direction and that the 2DES will form at the ETO/LAO interface. We chose a different material, namely the ferrimagnetic Mott insulator \GTO{} (GTO)~\rev{KomarekPRB}. In stoichiometric GTO the oxidation state of Ti is \tit. Therefore, GTO is polar, and the 2DEL can be formed at the GTO/STO interface. Interface conductance was indeed found in this system~\rev{MoetakefPRX,JacksonPRB}. Also, previous research demonstrated signatures of magnetism such as hysteretic behaviour of the magnetoresistance  and anisotropic magnetoresistance~\rev{MoetakefPRX,JacksonPRB}, but no AHE was detected~\rev{JacksonPRB}. Films of GTO are not as easy to grow as, for instance, LAO, and in the spirit of delta-doping we decided to grow heterostructure with ultrathin GTO layers, in particular LAO(8)/GTO(2)/STO, with the numeral denoting the number of unit cells. Characterization by electron microscopy showed intermixing of La and Gd, leading to a structure \LGAO /\GLTO /STO (LGAO/GLTAO/STO). Still, this serves our purpose, as it constitutes a system where magnetic ions are placed at, or close to, the conducting interface. Although the structures did not show ferromagnetism, we are going to show that 2DES in this system exhibit gate-tunable AHE, but only at low temperatures, where Rashba spin-splitting is essential. In particular, we find that the AHE at 3~K develops around a positive gate voltage of about 50~V, where the system passes through the Lifshitz transition. \\
\begin{figure*}
\includegraphics[scale=0.37]{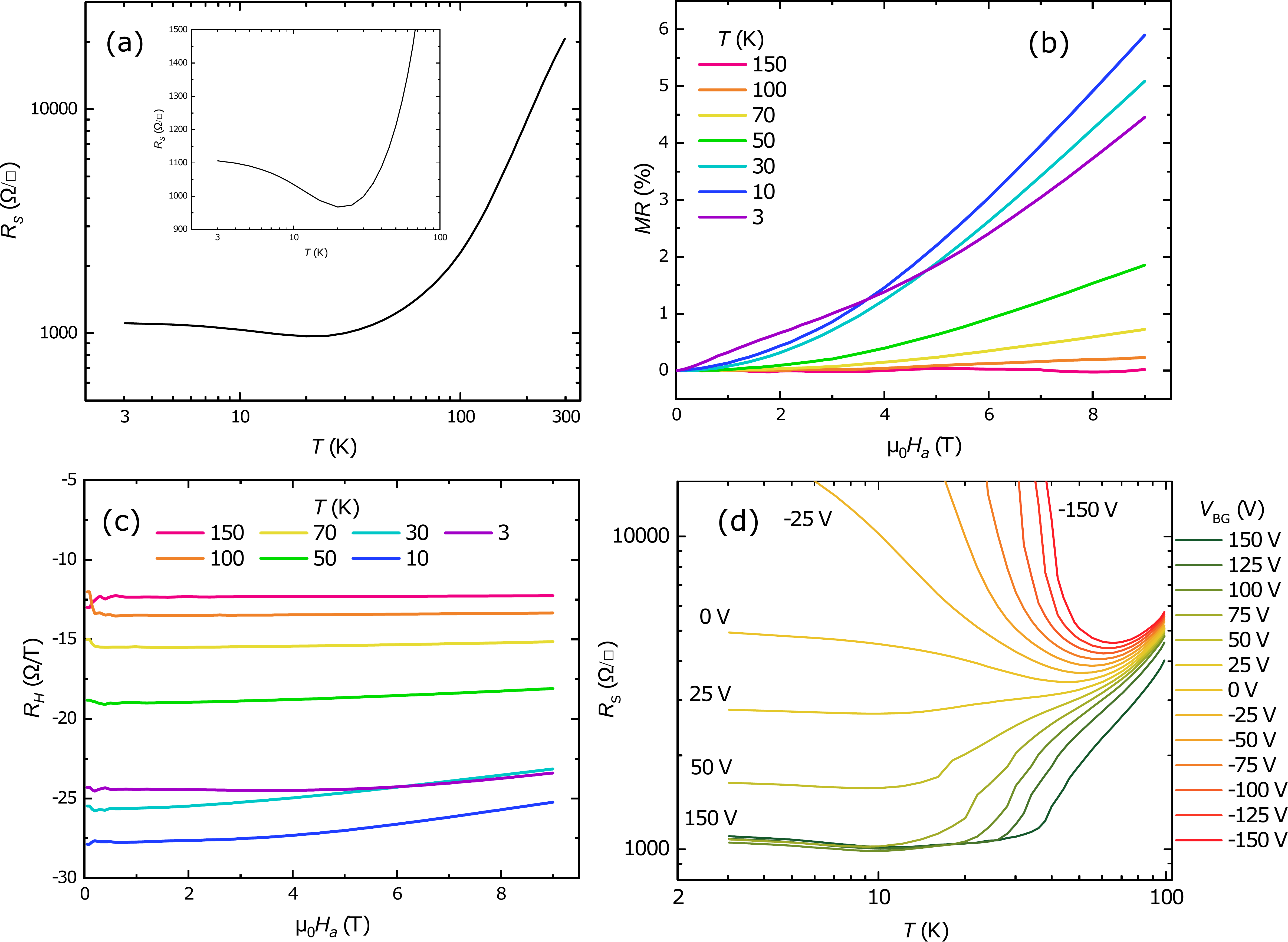}
\caption{(a) Temperature dependence of the sheet resistance $R_S$ of the LGAO/GLTAO/STO sample during cool down. The same data with linear Sheet resistance scale in lower temperature region.  (b) Field dependence of the magnetoresistance $MR$ and (c) field dependence of the Hall coefficient $R_H$. (d) Sheet resistance $R_S$ versus temperature at different back gate voltages $V_{BG}$ in the range from 150~V to -150~V as indicated.\label{Fig2}}
\end{figure*}
\section{Experimental Details\label{EXPER}}
Most of the previous studies on GTO heterostructures were performed on the films grown by Molecular Beam Epitaxy (see for example Ref.[\onlinecite{MoetakefJCG,MoetakefAPL,MoetakefAPL2}]) and only few by Pulsed Laser Deposition (PLD)~\rev{GrisoliaNPhys,GrisoliaAPL}.
In this work we grow samples on TiO\Ss{2}-terminated STO by PLD from an oxygen rich target \GTOx{} at 850 \DC{} at $1 \mx 10^{-4}$ mbar \ox{} nominal pressure. The repetition rate and laser fluency were set at 1 Hz and 1.3 J/cm\Us{2}, respectively. The samples were cooled down to room temperature at the growth pressure. The growth of the films was monitored by Reflection High-Energy Electron Diffraction (RHEED), which also yielded an estimate for the film thickness. Nominal layer thicknesses were chosen as 8 unit cell (u.c.) for LAO and ~2-2.5 u.c. for GTO. As will be shown later the real thicknesses and compositions of layers were different, due to strong intermixing. Also, the growth of a GTO layer with a reliable thickness turned out to be a challenging task due to the sensitivity to the growth conditions. In particular, rare earth titanates have a tendency to form a pyrochlore phase \PyroCl{} in an oxygen-rich environment~\rev{ScheidererAM,MoetakefAPL2}.
RHEED oscillations were hardly pronounced at the pressure we used, although the RHEED patterns did exhibit 2D growth (See Fig.~1 in the Supplementary information). At the same time, lowering of the \ox{} pressure would lead to enhancement of oxygen vacancies in STO and, therefore, to the bulk conductance in STO~\rev{KalabukhovPRB}. \\
Magnetotransport properties were measured using an automated measurement platform (a PPMS from Quantum Design) with a home built electrical insert to be able to gate the samples. They were measured in the van der Pauw geometry~\rev{VDP1,VDP2} at temperatures down to 3 K and magnetic fields up to 9 T. All measured field dependencies were (anti-)symmetrised. To study magnetism, scanning SQUID microscopy measurements were performed at 4.2 K without external magnetic field. A control sample showed the same qualitative behaviour as the results reported here. A second control sample was cut in two, with one part being used for analysis of the structure and chemical composition by Scanning Transmission Electron Microscopy (STEM) and  Electron Energy Loss Spectroscopy (EELS). The other part was used for transport measurements and showed results which were consistent with the earlier two samples.
\section{Results\label{Results}}
\subsection{TEM characterisation}
Analysis by STEM revealed that the films are crystalline (Fig. \ref{Fig1}a). At the same time EELS analysis indicated a severe intermixing in the sample (Fig. \ref{Fig1}b). In particular strong interdiffusion of Gd, La and Al is present over the whole thickness of the deposited layers, turning them into \LGAO{} and \GLTO{} instead of LAO and GTO respectively. Sr diffuses about 1~nm into the film whereas Ti diffuses further (around 2~nm)(Fig.~\ref{Fig1}b) yielding a thickness of GLTAO layer of about 5 u.c.
Further investigation of the structure of the \GLTO{} layer showed a varying amount of \tit{} and \tif{} along the film, as revealed by the study of the fine structure of the Ti L edges shown in figures \ref{Fig1}c-f. In the first 'GLTAO' region shown in Fig.~\ref{Fig1}c and d, Ti is purely in the \tit{} state (black, light blue and red spectra in Fig.~\ref{Fig1}d), whereas in the second 'GLTAO' region in Fig. \ref{Fig1}e and f the most of Ti is \tif. Data on the O K edge are shown in the supplementary information. Clearly, in spite of the capping with LAO, which should enhance the concentration of \tit, the growth of films in \ox{} atmosphere as well as the choice of STO as a substrate increase the concentration of \tif{} in the \RTO{} layer\rev{BibesDTO,ScheidererAM}.

\subsection{Basic transport properties\label{BTR}}
LGAO/GLTAO/STO is conducting and exhibit temperature dependence of sheet resistance($R_S$), which is comparable to LAO/STO\rev{BrinkmanNMAT} (Fig.~\ref{Fig2}a). Also the magnetoresistance $MR$ and the Hall resistance $R_{xy}$ were measured during cooldown. The magnetoresistance ($MR$) was calculated as:
\begin{equation}\label{eq1}
MR = \frac{R_S(B)-R_S(0)}{R_S(0)} \cdot 100\% .\\
\end{equation}
As shown in Fig.~\ref{Fig2}b, the $MR$ changes shape from almost flat to parabolic with decreasing temperature. At 3~K we note a different shape with a rather sharp dip around zero field, which indicates the appearance of Weak Anti-localization (WAL), similar to what was shown earlier for LAO/ETO/STO~\rev{PRBWLETO} and LAO/STO~\rev{CavigliaPRL, PRBWLLAO}. The Hall coefficient $R_H$ was extracted by dividing the Hall resistance by the applied field, $R_H = R_{xy} / (\mu_0 H_a)$. As shown in Fig. \ref{Fig2}c, $R_H$ starts to deviate from flat behavior (meaning a Hall resistance linear in the applied field) below 70 K. Such non-linear behaviour signals the presence of highly mobile \dxyz{} carriers\rev{JoshuaNCOM, SanderPRL}. At 3~K, a second non-linearity occurs at lower fields, in which the slightly parabolic shape around zero field becomes inverted. Such behavior has already been observed in other STO-based heterostructures and was identified as a signature of AHE~\rev{JoshuaNCOM, GunkelPRX, NMATLAOETO}. All in all, the basic transport characteristics show a behavior which is quite typical for that of the LAO/STO family.

\begin{figure*}[!]
\includegraphics[scale=0.28]{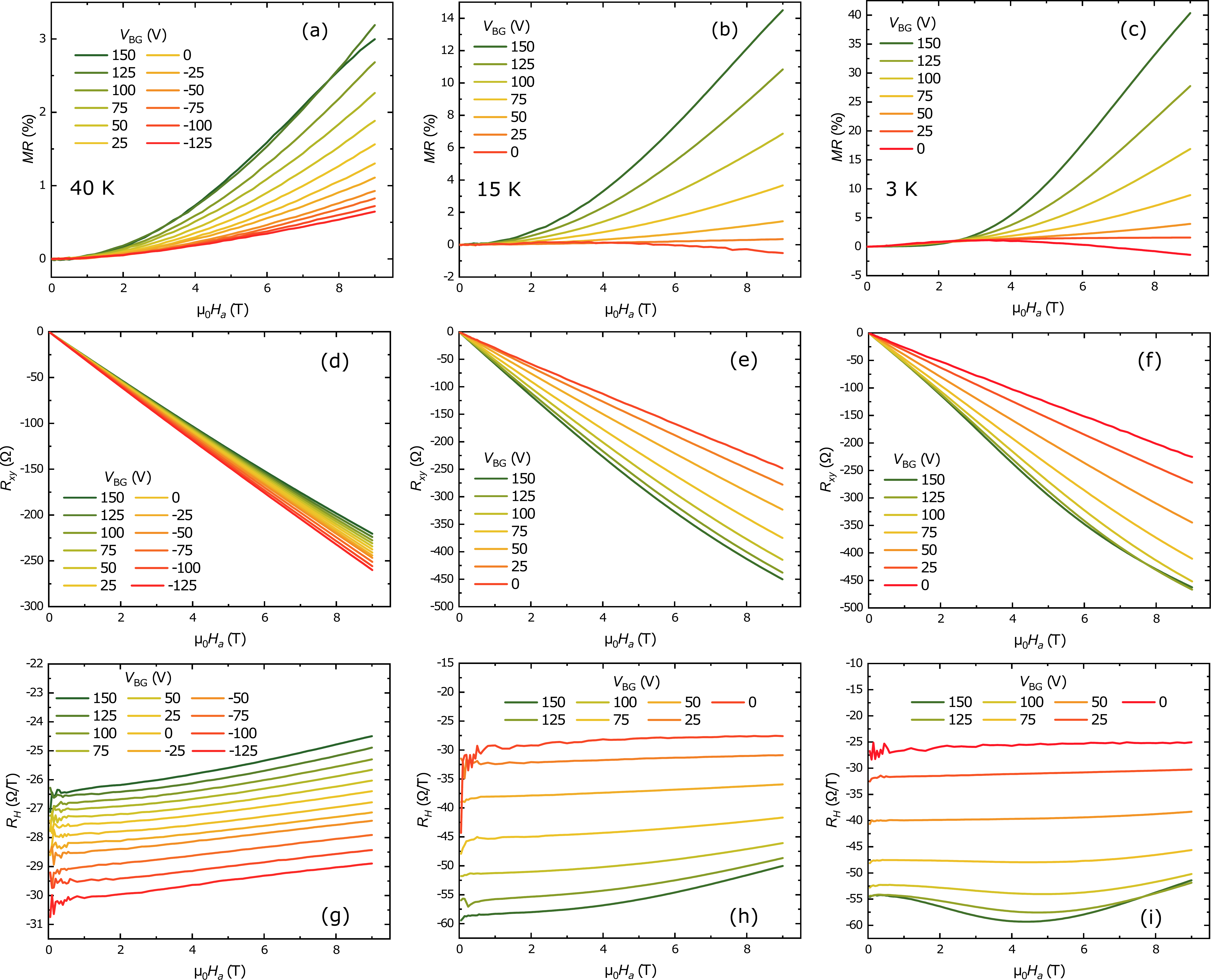}
\caption{Magnetoresistance $MR$, Hall resistance $R_{xy}$ and Hall coefficient $R_H$ of the LGAO/GLTAO/STO sample for different positive back gate voltages $V_{BG}$ as indicated at the temperatures of 40~K (a,d,g), 15~K (b,e,h) and 3~K (c,f,i). \label{Fig3}}
\end{figure*}

\begin{figure}
\includegraphics[scale=0.33]{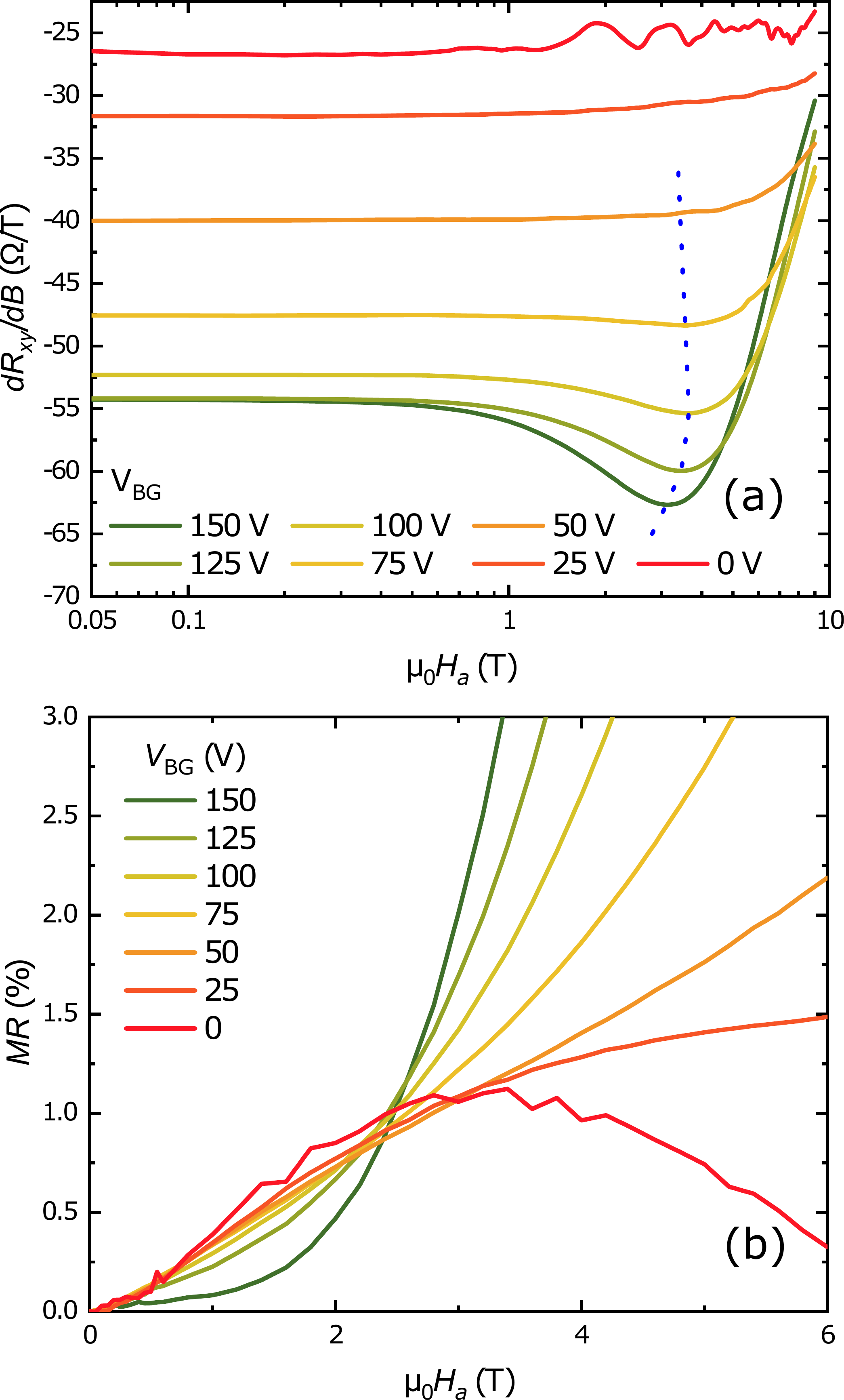}
\caption{(a) The derivative of the Hall resistance $R_H$ with respect to applied magnetic field $H_a$ (logarithmic scale), $dR_{xy}/d(\mu_0 H_a)$, at different positive gate voltages $V_{BG}$, for the LGAO/GLTAO/STO sample. The dotted line is to guide the eye. (b) Magnetoresistance $MR$ for different $V_{BG}$ at 3 K. The data are the same as in Fig.~\ref{Fig3}c, zoomed-in around the axes origin.\label{Fig4}}
\end{figure}

\begin{figure}[b]
\includegraphics[scale=0.33]{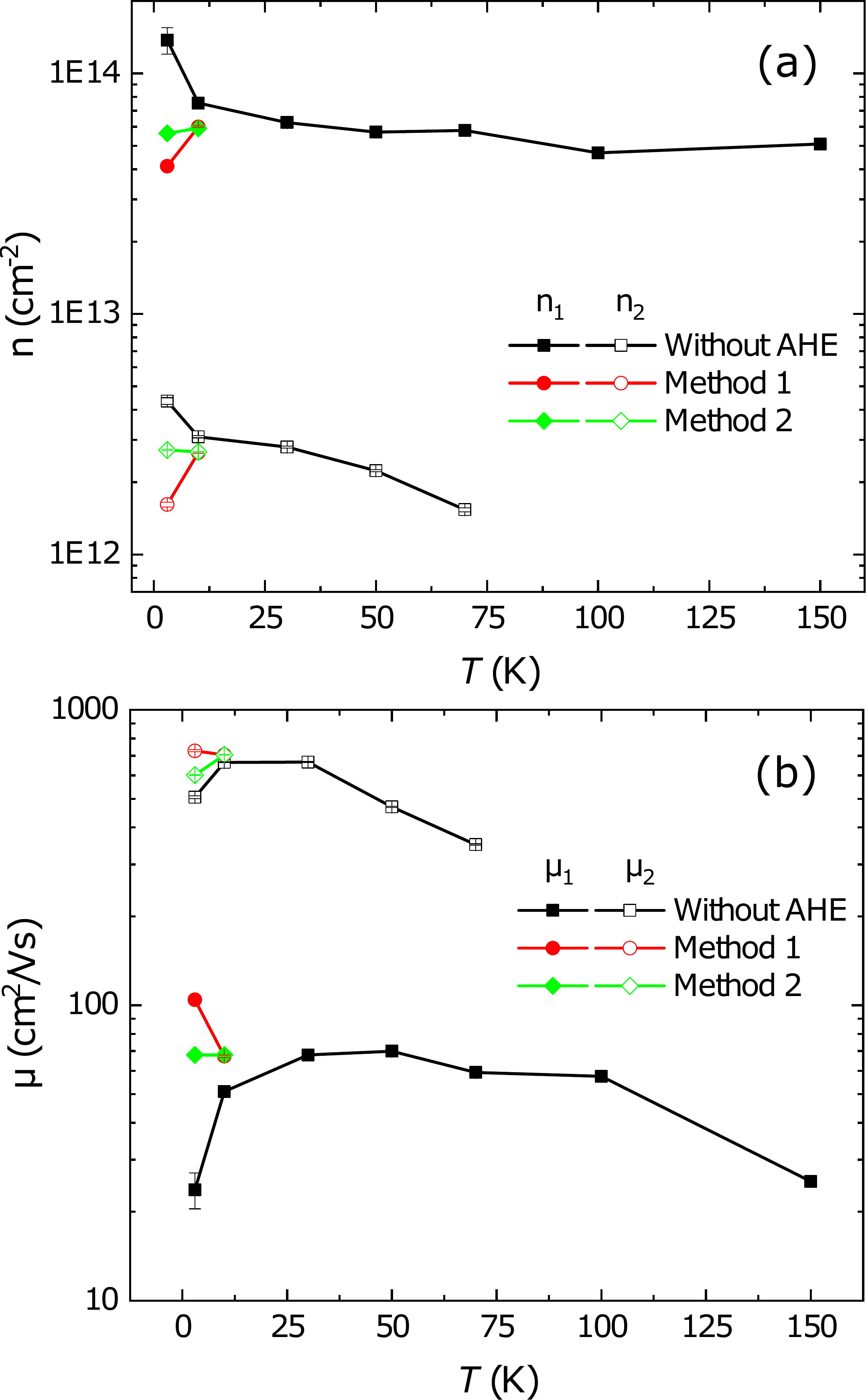}
\caption{(a) Carrier concentration $n$ and (b) mobility $\mu$ versus temperature for the LGAO/GLTAO/STO sample.\label{Fig5}}
\end{figure}

\begin{figure*}[!]
\includegraphics[scale=0.28]{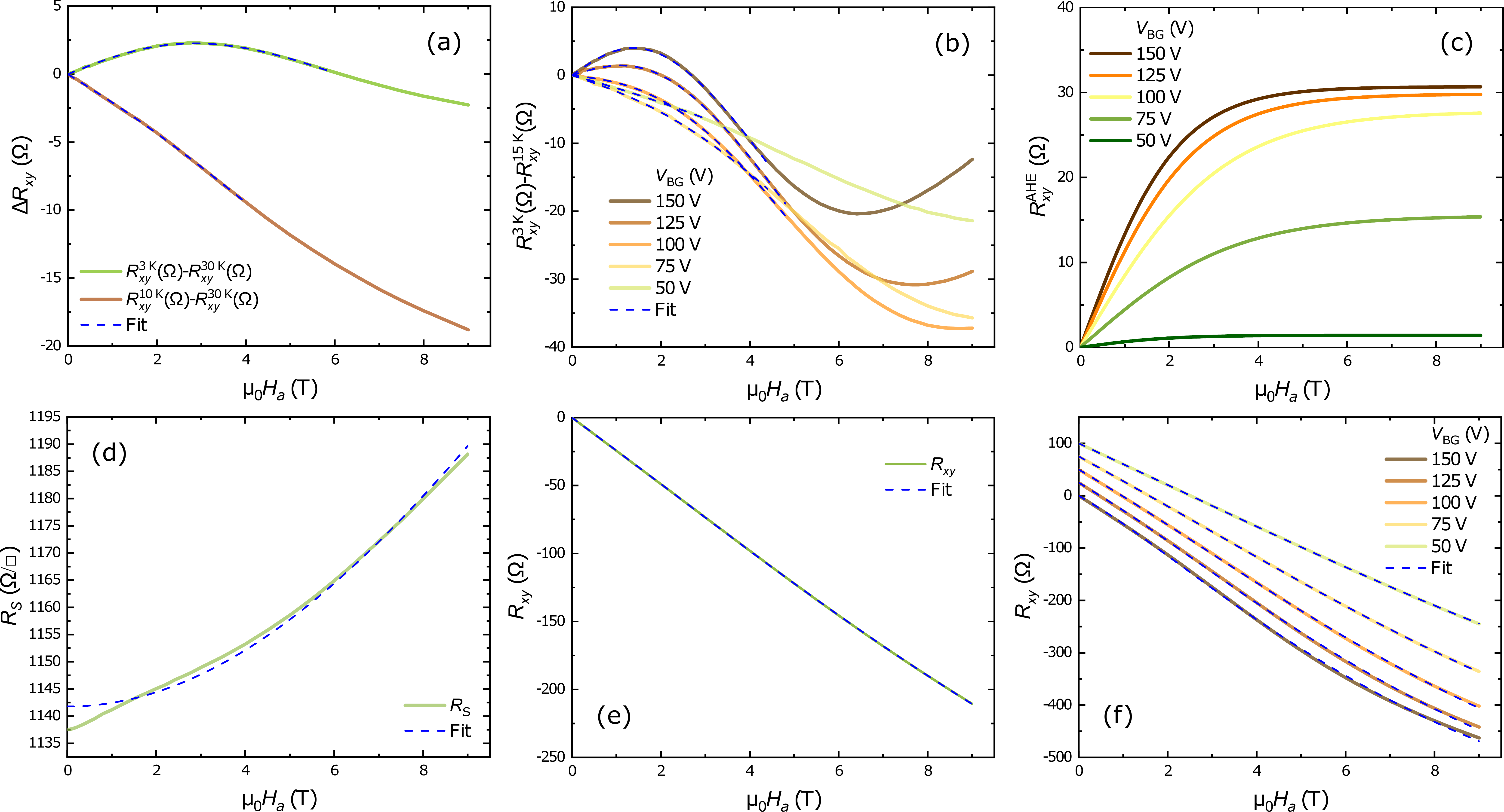}
\caption{(a) Result of the substraction of the Hall resistance $R_{xy}^{10K}$ measured at 10 K from the curve measured at 3 K before applying a back gate voltage $V_{BG}$. (b) Substraction of the curves measured at at 15 K and $V_{BG}$ from curves measured at 3 K (solid lines) and fit with Eq.~\ref{eq10}(dotted lines). (c) Resulting AHE field dependence obtained from the fit by Method 1 (see text). Fit of (d) sheet resistance $R_S$ and (e) Hall resistance $R_{xy}$ by Method 2 at 3~K before applying the gate voltage. (f) Fit of $R_{xy}$ at different $V_{BG}$. An offset of 25 $\Omega$ between curves was added for clarity.\label{Fig6}}
\end{figure*}

\begin{figure}
\includegraphics[scale=0.33]{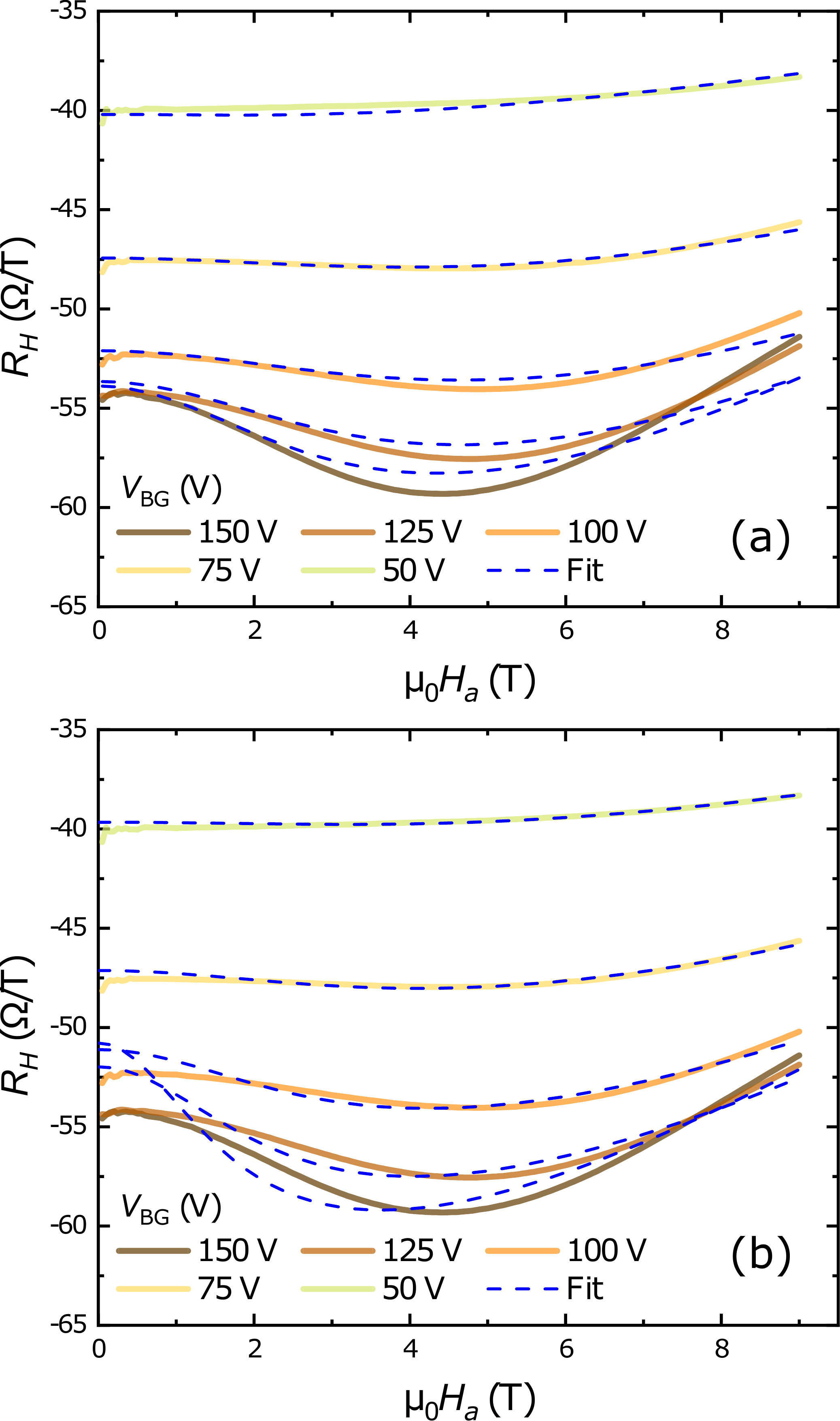}
\caption{Hall coefficient $R_H$ at 3 K and different gate voltages $V_{BG}$ (solid lines) and fit by (a) Method 1 and (b) Method 2 (dotted lines).\label{Fig7}}
\end{figure}

\subsection{Effects of gating}
Next, we studied the behavior of the LGAO/GLTAO/STO sample upon applying a back gate voltage $V_{BG}$. Gating results in a tunable Metal-Insulator Transition (MIT), as shown in Fig. \ref{Fig2}d. The $R_S(T)$ curves were measured by cooling down from 100~K at constant applied gate voltage. The change of the gate voltage, going down from 150~V, was always performed at 3~K. Below -25~V, the system becomes insulating at low temperatures. We do not observe saturation of $R_S$ in our samples,  similar to delta-doped samples with SrTi\Ss{1-x}Mn\Ss{x}O\Ss{3}\rev{FixAPL} but in contrast to such materials as ETO and \LCO, where a Kondo-like effect with saturation at low temperatures\rev{LucaPRB, DasPRB} was observed. Furthermore, the system shows incipient Weak Localization (WL) behavior (see below), which also has been observed in LAO/STO\rev{CavigliaPRL,LiaoPRB,PRBWLLAO}. In the range from -25~V to 0, a pronounced minimum appears as function of temperature. For positive gate voltages, a small upturn in $R_S$ can be observed. Ref. [\onlinecite{GunkelPRX}] indicated a correlation of a similar upturn and the emergence of AHE in a similar 2DES system, \NGO/STO (NGO/STO). At the same time, the MR changes shape from incipient WL to WAL, which disappears at high positive $V_{BG}$. All data confirm that the behaviour of LGAO/GLTAO/STO follows the scenario well known for LAO / STO, with the presence of a localised phase at negative $V_{BG}$, and a cross-over to a conducting phase at positive $V_{BG}$, ascribed to  a strong change in (Rashba) spin-orbit interaction by Caviglia {\it et al.}\rev{CavigliaPRL}.\\

To better understand the transport behaviour at low temperatures, and in particular at the lowest temperature of 3~K, we studied the evolution of the magnetotransport properties as function of back gate voltage more closely at three different temperatures, 40~K, 15~K and 3 K. All gate voltage sweeps were made by sweeping from 150~V to downward, ending at 0~V for 15~K and 3~K because of the MI transition. Comparing the results shows significant changes occurring when going to the lowest temperature. At 40~K the $MR$ is small and has a parabolic shape in the whole range of gate voltages (Fig. \ref{Fig3}a). At 15 K (Fig. \ref{Fig3}b) and 3 K (Fig. \ref{Fig3}c) the shape, in particular at low $V_{BG}$ diverges from parabolic, and a negative $MR$ appears in high field. The (negative) Hall resistance at 40~K decreases with decreasing gate voltage, but increases at 15~K and 3~K.(Fig. \ref{Fig3}d-f). The Hall coefficient $R_H$ shows that the
Hall effect is non-linear in the whole range of voltages at 40~K (Fig. \ref{Fig3}g). At 15 K it becomes non-linear above 25~V and at 3~K  above 50~V (Fig. \ref{Fig3}h,i). As was mentioned above, this non-linearity of the Hall effect in high fields indicates the presence of \dxyz{} carriers, and their appearance signals that the system passes through the Lifshitz point. This is of obvious importance for the extraction of carrier concentrations and mobilities~\rev{JoshuaNCOM, SanderPRL}.

\subsection{The question of ferromagnetism}
In the previous section we found, below 70~K and above the gate-induced Lifshitz transition, a characteristic dip structure in the field dependence of $R_H$ (Fig.~\ref{Fig3}i), which indicates the presence of the AHE. Appearance of this dip above 50 V can be better seen in $dR_{xy}/dB$ field dependence plotted on Fig. \ref{Fig4}a). The observation confirms the crucial role of \dxyz{} carriers in the AHE and is in agreement with previous results on LAO/ETO/STO\rev{NMATLAOETO}. However, we did not observe AHE at 15 K and 40 K in spite of the clear presence of the second type of carriers (Fig. \ref{Fig3}g and h). At these temperatures signatures of WAL were completely absent, indicating a possible important role for SOC in observation of AHE. Although the onset and increase of the AHE with increasing gate voltages is accompanied by the disappearance of WAL (Fig. \ref{Fig4}b), Stornaiuolo and co-workers argued that the appearing of AHE may mask spin-orbit coupling rather than suppress it\rev{PRBWLETO}. Moreover, with increasing gate voltage the carrier concentration is also increasing, leading to stronger contribution of orbital effects in out of plane MR. \\
The occurrence of AHE is often taken as a signature of ferromagnetism, but we were not able to detect any hysteresis in our magnetotransport measurements down to 3~K. The same behaviour was reported for an interface between paramagnetic NGO and STO \rev{GunkelPRX}. The occurrence of AHE was explained with the polarization of magnetic moments, more specifically by the rotation of moments around the out-of-plane hard axis in a magnetic field perpendicular to the sample surface\rev{LiNPHYS}. In order to investigate this further, we performed measurements with a scanning SQUID microscope~\rev{PimSQUID} on non-gated samples and samples gated at 150 V, 0 V and -150 V. The resulting scans did not show any signatures of ferromagnetic domains at 4.2~K, nor did they show ferromagnetic patches such as observed by scanning SQUID in LAO/STO structures~\rev{KaliskyNCOM,KaliskyNPHYS,KaliskyNL}. This could be due to the fact that domains are smaller than our resolution. At the same time, however, EELS data indicated the presence of regions with \tit and regions without it, and such a distribution of \tit can lead to superparamagnetic behaviour. The measurement setup did not allow to apply a magnetic field and gate voltage simultaneously so we cannot completely exclude a scenario of superparamagnetic rather than paramagnetic behaviour, in which larger ferromagnetic domains form in an external magnetic field. Also, the absence of a change in the ferromagnetic landscape when applying a gate voltage is consistent with study on LAO/STO~\rev{KaliskyNCOM}. The premise for the remainder of the paper therefore is that the magnetic Ti- and Gd-moments which are present, are polarizable but not ordered. The question to be answered is whether a meaningful AHE contribution can be extracted from the data, and then how to extract meaningful numbers for the carrier concentrations and mobilities, using Hall effect measurements in which AHE are present. \\

\subsection{Extracting the Anomalous Hall Effect\label{EAHE}}
Firstly,we calculate the longitudinal $G_{xx}$ and transverse $G_{xy}$ conductance from the relationship between resistance and conductance.
\begin{equation}\label{eq2}
G_{xx}(B)= \frac{R_{S}(B)}{R_{S}^2(B)+R_{xy}^2(B)},\\
\end{equation}
\begin{equation}\label{eq3}
G_{xy}(B)= \frac{-R_{xy}(B)}{R_{S}^2(B)+R_{xy}^2(B)}, \\
\end{equation}

Then, to extract carrier concentrations and mobilities, we fit both of them using a two-band model.
\begin{equation}\label{eq4}
G_{xx}(B)= \frac{en_1\mu_1}{1+\mu_1^2B^2}+\frac{en_2\mu_2}{1+\mu_2^2B^2}, \\
\end{equation}
\begin{equation}\label{eq5}
G_{xy}(B)= \frac{en_1\mu_1^2B}{1+\mu_1^2B^2}+\frac{en_2\mu_2^2B}{1+\mu_2^2B^2}, \\
\end{equation}
where $e$ is the elementary charge. The temperature dependencies of both quantities is shown in Fig.\ref{Fig5}~and yields the usual picture. We find a high-mobility band with low carrier concentration (of order $3 \times 10^{12}$~cm$^{-2}$) and a low-mobility band with high carrier concentration (of order $8 \times 10^{13}$~cm$^{-2}$). The data at 3~K show sharp changes in all values, however, because the model does not capture the contribution of the AHE to the Hall data. Gunkel et al.\rev{GunkelPRX} showed that this additional contribution can be described as the behaviour of a superparmagnet, which essentially is the Langevin function for the paramagnetic behavior of a cluster of magnetic moments~:
\begin{equation}\label{eq6}
R^{AN}(B)= R_{xy}^{AHE} tanh\left(\frac{B}{B_c}\right). \\
\end{equation}
Here \RAN{} is the Anomalous Hall coefficient and B\Ss{c} is another fitting parameter which takes over the role of temperature in the original Langevin description in setting the energy scale for the field. \\

In a different approach, Maryenko et al.~\rev{MaryenkoNCOMM} used a Brillouin function to describe the non-hysteretic Anomalous Hall effect in a non-magnetic 2DEG based on the very different system MgZnO/ZnO~:
\begin{equation}\label{eq7}
B_J(x)= \frac{J+1}{2J}coth\left(\frac{J+1}{2J}x\right)-\frac{1}{2J}coth\left(\frac{1}{2J}x\right),  \\
\end{equation}
where
\begin{equation}\label{eq8}
x= \frac{gm_{eff}\mu_B JB}{k_B T},  \\
\end{equation}
$g$ is the $g$-factor, $J$ is the total orbital angular momentum, and $m_{eff}$ is the effective magnetic moment averaged over the whole sample in units of the Bohr magneton ($\mu_B$). In the case of $g=2$ and $J=1/2$ eq.~\ref{eq7} reverts to eq.~\ref{eq6}~:
\begin{equation}\label{eq9}
R^{AN}(B)=R_{xy}^{AHE} tanh\left(\frac{m_{eff}\mu_B B}{k_B T}\right), \\
\end{equation}
with $M_{eff}$ and \RAN{} both fitting parameters. \\
Based on this last function we implemented two ways to extract the carrier concentrations at 3~K. The first one is a "subtraction" method or method~1 later in the text. The idea is simply that in a small range of temperatures the mobilities and concentrations would not change abruptly or too much, and it will be possible to find the AHE by subtracting the Hall resistance curve at higher temperature without AHE from the Hall resistance curve with AHE at lower temperature. Fig.\ref{Fig6}a shows the result of subtracting $R_{xy}$(30~K) from $R_{xy}(3 K)$ and $R_{xy}(10 K)$ before applying a back gate voltage. As can be seen from Fig.\ref{Fig6}a the AHE contribution at 3~K is clearly visible as a Brillouin-like function. At 10K ~the contribution is much smaller. Also, the slope of the residual linear contribution is negative for the 10~K - 30~K subtraction, as a consequence of the behavior of $R_H$ shown in Fig.~\ref{Fig2}c where the 10~K curve is shifted upward with respect to the 30~K curve.  The AHE contribution at 3~K can be fitted quite well with Eq.~\ref{eq9} when we add a linear residual low field ordinary Hall resistance :
\begin{equation}\label{eq10}
R^{AN}(B)=R_{xy}^{AHE} tanh(\frac{M_{eff}\mu_B B}{k_B T})+a \; B, \\
\end{equation}
where $a$ is the slope of residual Ordinary Hall Resistance in low field and one of the fitting parameters, along with $R_{xy}^{AHE}$ and $M_{eff}$. The fit yields \RAN =9.3~$\Omega$ and $m_{eff}$=1.33 for 3 K and  \RAN =1.78~$\Omega$ and $M_{eff}$=6.03 for 10 K. Fitting the  Brillouin function (\ref{eq7}) with higher $J$ does not improve the fit. In a similar way we subtracted curves at 15 K from 3 K measured at different gate voltages (Fig. \ref{Fig6}b). In this case the residual linear contribution is negative because of the increase in carrier concentration, especially of the high mobility type, after passing through the Lifshitz transition. Fig.~S3 in the supplementary information shows the result of subtraction of curves at 40 and 15 K at different gate voltages. In that case only a high field non-linearity is present. Fig.~\ref{Fig6}c shows the curves as they came out from fitting Eq.~\ref{eq9}. These curves were then subtracted from $R_{xy}$ in order to be able to use Eq.~\ref{eq4} and Eq.~\ref{eq5} and obtain carrier concentrations and mobilities. \\

\begin{figure*}[!]
\includegraphics[scale=0.42]{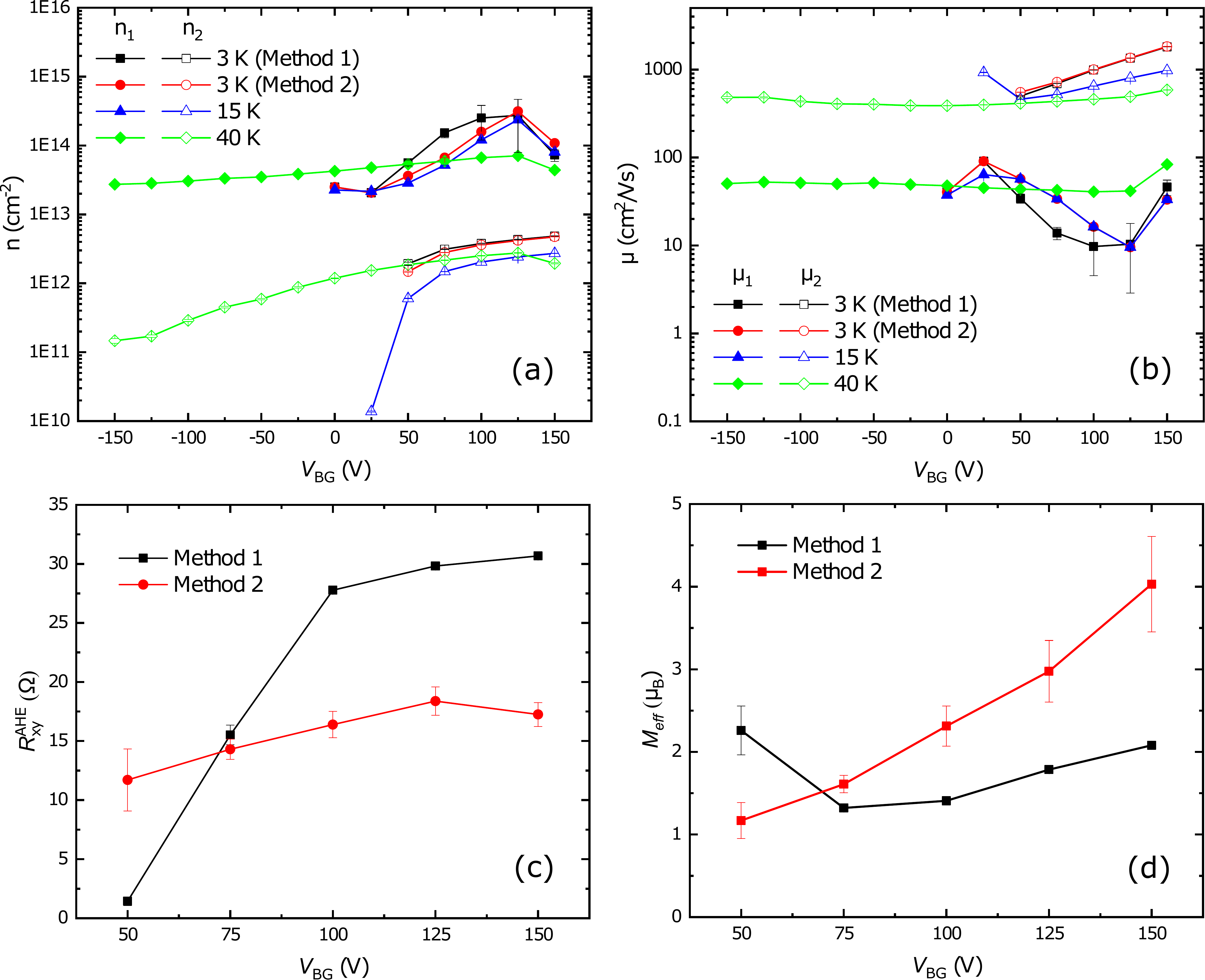}
\caption{(a) Carrier concentration, and (b) mobility as function of gate voltage for different temperatures. (c) Anomalous Hall coefficient and (d) magnetic moment versus backgate voltages at 3 K.\label{Fig8}}
\end{figure*}

The second method (method 2) is based on directly fitting $R_{S}$ and $R_{xy}$, using a model similar to the one used by Gunkel et al.\rev{GunkelPRX}:
\begin{equation}\label{eq11}
R_{S}(B)= \frac{G_{xx}(B)}{G_{xx}^2(B)+G_{xy}^2(B)},\\
\end{equation}
\begin{equation}\label{eq12}
R_{xy}(B)= \frac{-G_{xy}(B)}{G_{xx}^2(B)+G_{xy}^2(B)}+R^{AN}(B), \\
\end{equation}
where $R^{AN}(B)$ is in form of Eq.~\ref{eq9}. To make the fit converge, the mobility of the low mobility carriers was fixed at the value obtained from the fit at 15~K. Fitting of the curves before applying the gate voltage gives a result quite close to that of  Method 1: \RAN =4.36 $\Omega$ and $m_{eff}$=1.78 for 3~K and \RAN =3.13 $\Omega$ and $m_{eff}$=4.41 for 10~K. \\
The fitted curves for the initial cool down are shown for $R_S$ in Fig.~\ref{Fig6}d and for $R_{xy}$ in Fig.~\ref{Fig6}e.
The fit of the sheet resistance is not able to catch the WAL behavior at low field (Fig. \ref{Fig6}d), but otherwise it works well.
The resulting fits for the Hall resistance are displayed in Fig.~\ref{Fig6}f. \\
Both methods remove the sharp increase of carrier concentrations and decrease of mobilities at 3~K (Fig.~\ref{Fig5}). However, a closer look at the Hall coefficient (Fig. \ref{Fig7}), which is a more sensitive parameter than $R_{xy}$ itself, reveals that the first method describes the low field dependence better, while the second one is better for the high field dependence. The resulting fitting parameters $n, \mu, R_{xy}^{AHE}$ and $M_{eff}$ for both methods are presented in Fig.~\ref{Fig8}, which can be considered as one of the main results of the paper. \\
We can note a few things. At 40~K a value for the carrier concentration of the second band is found for all gate voltages (Fig. \ref{Fig8}a). For 15~K and 3~K, those carriers are only found for positive gate voltage. Carriers from the first band show a significant increase in concentration and a significant decrease in mobility upon gating at 15 and 3 K (Fig. \ref{Fig8}a and b). Generally, the results of both methods to extract carrier concentration and mobility in the presence of AHE do not show significant differences, especially for the high mobility carriers. \RAN{} extracted by method 1 shows more increase, whereas the results obtained by method 2 are less clear due to larger error bars(Fig. \ref{Fig8}c). The results of method one show saturation behavior, similar to what was found for LAO/ETO/STO~\rev{NMATLAOETO}{}. The magnetic moment ($M_{eff}$) shows an increase with increasing gate voltage(Fig. \ref{Fig8}d), which is possibly linked to the increasing carrier concentration of the low mobility band.

\section{Discussion\label{Discussion}}
We have shown in the previous sections that, in particular at 3~K, a non-hysteretic AHE contribution to the Hall resistance can be found. Here we discuss its possible origins. Naively, the answer could be that ferromagnetism is induced by the insertion of ferrimagnetic GTO. We were not able to detect such regions by scanning SQUID microscopy. Taking into account results of EELS and scanning SQUID analysis, we conclude that our GLTAO layer is rather in a paramagnetic or superparamagnetic state than ferrimagnetic, due to strong intermixing. \\
Strictly speaking, AHE need not be signature of ferromagnetic order. A growing number of reports, both theoretical and experimental, shows that it can be seen in other systems, for example in paramagnets\rev{MaryenkoNCOMM, CumingsPRL, CulcerPRB,CuclerPRB2, FertPRB, ChazalvielPRL}, in superparamagnets\rev{ShindePRL,YANGJAC,ZhangPRB}, in antiferromagnets\rev{NakatsujiNat,ChenPRL} and, as already mentioned in \LAO/\STO{} through magnetism at ferroelastic domain walls in the STO\cite{ChrisNPhys}.
%
%
In our case, the AHE increases with gate voltage and therefore appears to be connected to the transition through the Lifshitz point and the onset of conductance through a second band. This allows for two new mechanisms to appear. One has to do with the magnetic interactions. The \dxyo{} band is circular, lies in the plane of interface, and \dxy{} electrons are coupled antiferromagnetically to magnetic moments\rev{RuhmanPRB,JoshuaPNAS}. The \dxyzo{} bands, on the other hand, have highly elliptical Fermi surfaces directed along crystal axes and the \dxyz{} electrons are thought to couple ferromagnetically to the local Ti$^{3+}$ magnetic moments\rev{RuhmanPRB,JoshuaPNAS}. Ferromagnetic interactions therefore may appear beyond the Lifshitz point. The other and probably more important mechanism is the enhanced Rashba spin splitting occurring near the band crossing between the light and heavy bands\rev{KingNCOMM,SeilerPRB}, which leads to amplified spin-orbit coupling\rev{DiezPRL,BovenziPRB}. Specifically, it has been shown that the characteristic spin-orbit fields can increase almost an order of magnitude with increasing gate voltage \cite{BenShalomPRL,CavigliaPRL,LiangPRB}. Against this picture pleads that AHE in our system has been observed only at temperatures around or below 10 K, whereas the two band behaviour is generally observed in a much wider temperature range. However, the physics here is more complicated than in the case of semiconductors. Diez {\it et al}\rev{DiezPRL}, for instance, showed that, around the Lifshitz point, the density of states increases steeply with band energy which leads to a strong lowering of the chemical potential between about 5~K and 20~K. Such effects should also affect the magnetic interactions leading to AHE. All in all, there are ample indications that the gate-induced onset of AHE at the lowest temperatures is due to the physics of the Lifshitz-point, while the disappearance of AHE at higher temperature, notwithstanding the presence of two bands, could be explained by the same physics.

\section{Conclusion\label{Conclusion}}
In conclusion, we have observed the occurrence of the Anomalous Hall Effect in intermixed layers of LGAO/GLTAO/STO upon applying a positive back gate voltage at low temperatures, and without observing signs of ferromagnetism. We implemented an alternative method to the one of ref.[\onlinecite{GunkelPRX}] in order extract the AHE coefficient, as well the carrier concentrations and mobilities in the two band electron system. We pointed out that the onset of AHE is found at low temperatures when the system undergoes a Lifshitz Transition and  Rashba spin splitting is enhanced. The physics we observe appears to be quite robust, and relatively independent of the 2DES system being researched.

\begin{acknowledgments}
N.L. and J.A. gratefully acknowledge the financial support of the research program DESCO, which is financed by the Netherlands Organisation for Scientific Research (NWO). J.V. and N.G. acknowledge funding from the Geconcentreerde Onderzoekacties (GOA) project “Solarpaint” of the University of Antwerp and the European Union’s horizon 2020 research and innovation programme ESTEEM3 under grant agreement \textnumero 823717. The Qu-Ant-EM microscope used in this study was partly funded by the Hercules fund from the Flemish Government.
Authors thank A.E.M. Smink, L. Krusin-Elbaum, N. Bovenzi and G. Koster for useful discussion.
\end{acknowledgments}

\bibliography{GTO}
\end{document}


\newcommand{\Ss}{\textsubscript}
\newcommand{\Us}{\textsuperscript}
\newcommand{\STO}{SrTiO\Ss{3}}
\newcommand{\GTO}{GdTiO\Ss{3}}
\newcommand{\LAO}{LaAlO\Ss{3}}
\newcommand{\RTO}{ReTiO\Ss{3}}
\newcommand{\PyroCl}{Re\Ss{2}Ti\Ss{2}O\Ss{7}}
\newcommand{\ETO}{EuTiO\Ss{3}}
\newcommand{\ox}{O\Ss{2}}
\newcommand{\tit}{Ti\Us{3+}}
\newcommand{\tif}{Ti\Us{4+}}
\newcommand{\DC}{\degree C}
\newcommand{\mx}{\times}
\newcommand{\rev}{\cite}
\newcommand{\RAN}{R\Ss{xy}\Us{AN}}
\newcommand{\dxy}{3d\Ss{xy}}
\newcommand{\dxyz}{3d\Ss{xz/yz}}


\title{Supplementary Information for Gate-tuned Anomalous Hall Effect Driven by Rasba Splitting in Intermixed \LAO/\GTO/\STO}


\author{N. Lebedev}
\affiliation{Kamerlingh Onnes Laboratory, Leiden University, P.O. Box 9504, 2300 RA Leiden, The Netherlands}
\author{M. Stehno}
\affiliation{Physikalisches Institut (EP 3), Universität Würzburg, Am Hubland 97074 Würzburg, Germany}
\author{A. Rana}
\affiliation{Center for Advanced Materials and Devices, BML Munjal University (Hero Group), Gurgaon, India - 122413}
\affiliation{MESA+ Institute for Nanotechnology, University of Twente, P.O. Box 217, 7500 AE Enschede, The Netherlands}
\author{P. Reith}
\affiliation{MESA+ Institute for Nanotechnology, University of Twente, P.O. Box 217, 7500 AE Enschede, The Netherlands}
\author{N. Gauquelin}
\affiliation{Electron Microscopy for Materials Science, University of Antwerp, Campus Groenenborger Groenenborgerlaan 171, 2020 Antwerpen, Belgium}
\author{H. Hilgenkamp}
\author{A. Brinkman}
\affiliation{MESA+ Institute for Nanotechnology, University of Twente, P.O. Box 217, 7500 AE Enschede, The Netherlands}
\author{J. Aarts}
\affiliation{Kamerlingh Onnes Laboratory, Leiden University, P.O. Box 9504, 2300 RA Leiden, The Netherlands}

\date{\today}


\maketitle

\begin{figure}
\includegraphics[scale=0.33]{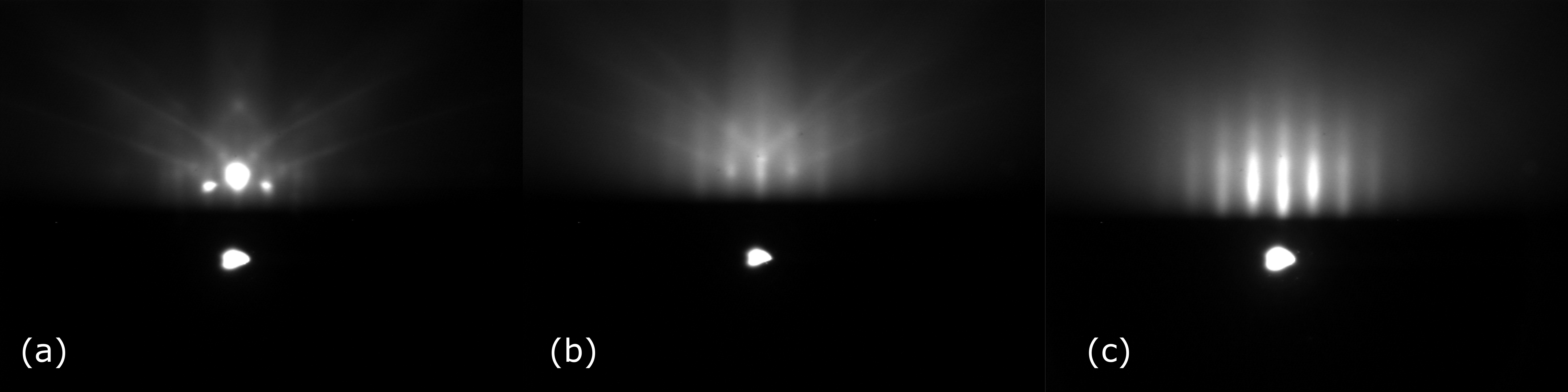}
\caption{RHEED patterns (a) before deposition at room temperature, (b) after deposition of \GTO{} and (c) after deposition of \LAO{} at 850\DC.\label{SupFig2}}
\end{figure}

\begin{figure}
\includegraphics[scale=0.24]{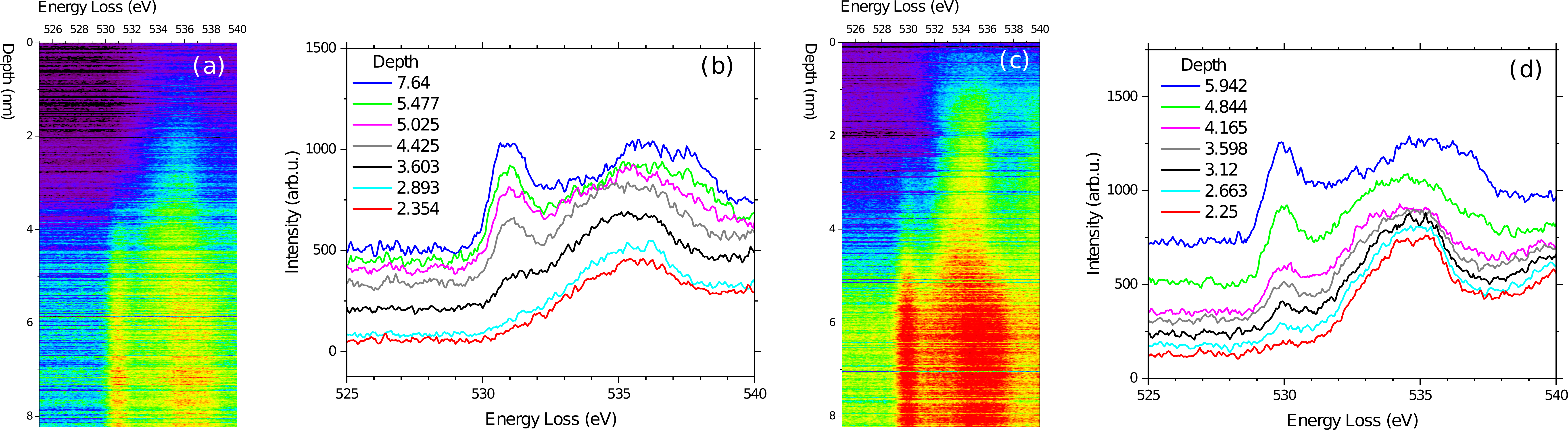}
\caption{EELS data; (a) O K-edge signal as function of distance from the interface in a region where \tit{} is present in the GLTAO layer and (b) corresponding spectra in that region.  (c) O K edge in a region where \tit{} is absent in the GLTAO layer and (d) corresponding spectra in that region.}
\end{figure}

\begin{figure}
\includegraphics[scale=0.6]{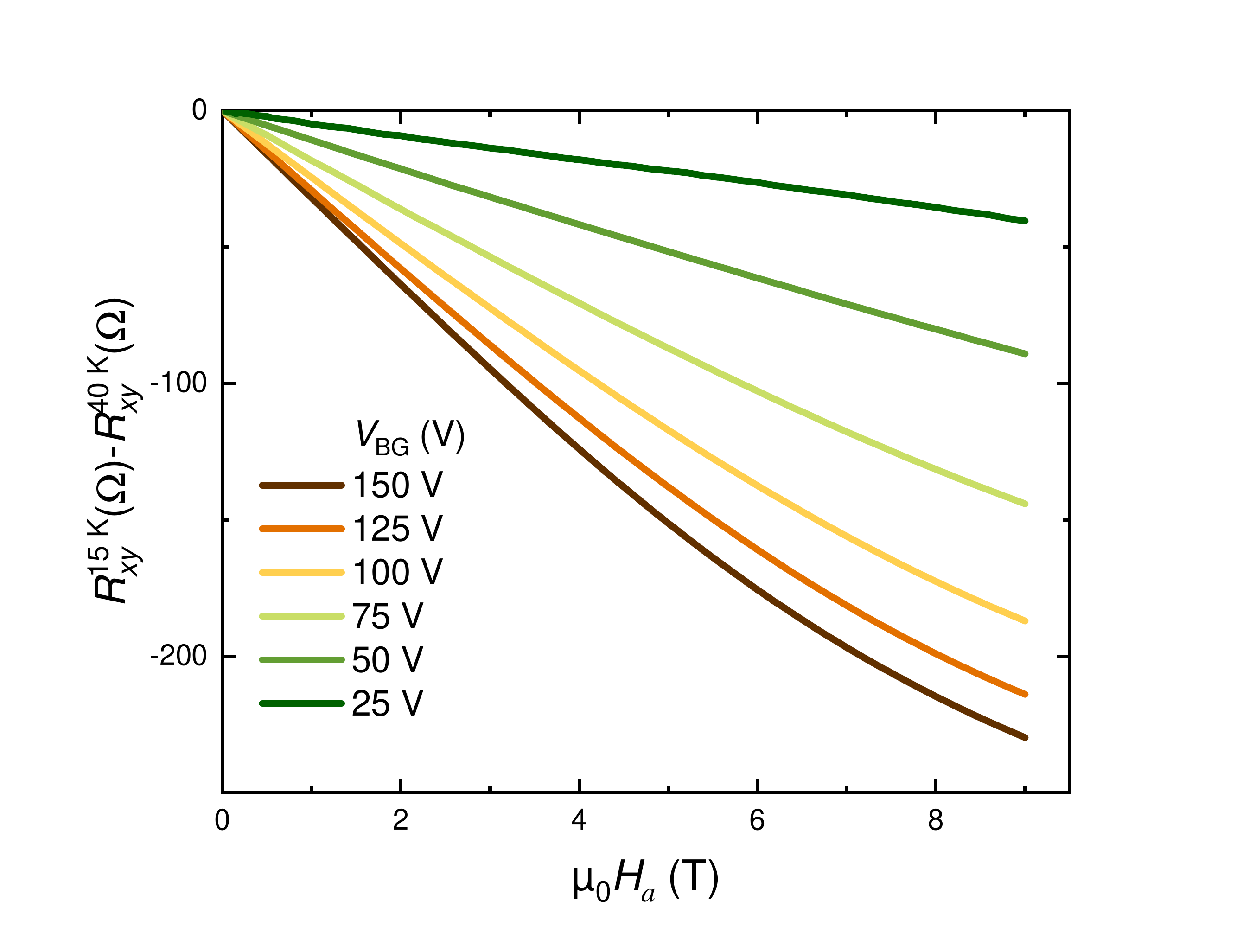}
\caption{The result of substraction of Hall resistances $R_{xy}^{40 K}$ measured at 40 K from curves measured at 15 K, at different gate voltages as indicated.}
\end{figure}